\documentstyle[icproc,epsf,psfig]{article}
\begin{document}



\newcommand{\ra}{\rightarrow}
\newcommand{\ko}{K^0}
\newcommand{\be}{\begin{equation}}
\newcommand{\ee}{\end{equation}}
\newcommand{\bea}{\begin{eqnarray}}
\newcommand{\eea}{\end{eqnarray}}

\lecture{(HALF) A LECTURE ON D-BRANES}

\author{ C. BACHAS
\thanks{Lecture at the Workshop on Gauge Theories, Applied Supersymmetry
and Quantum Gravity, Imperial College, London (July 1996), and
at the Institut d' Et\'e, Ecole Normale Sup\'erieure, Paris (August 1996).
 }  }

\address{Centre de Physique Th\'eorique, Ecole Polytechnique,\\
91128 Palaiseau, FRANCE\\
email: bachas@orphee.polytechnique.fr}

\maketitle
\abstracts{ This is a concise foreword to,  rather than a review of,
 D-brane physics.
}

\section{Introduction}
 
 In their early days dual models  were proposed as a theory
of hadronic interactions. Mesons are described by open
strings, and in this context it is natural  to 
consider heavy quarks at their endpoints. D-branes are generalized
heavy quarks,  which in the context of a fundamental theory
acquire, both poetically and literally, 
some completely novel dimensions. First, in contrast to
the quarks of QCD,  
D-branes are believed to be intrinsic to the fundamental theory:
without them there could be no web of dualities,~\cite{Dual,Duall}
 holding all
consistent superstrings inside a unique though still mysterious
(${\cal M}$ or ${\cal F}$?) structure.
Second, being non-perturbative excitations of gravity, 
D-branes are in the process of modifying drastically our thinking
about quantum space-time. The fascinating interplay of
supersymmetric gauge theory and geometry has shed some
surprising new light on both. Among all the exciting
recent developments, this could prove the lesson whose consequences
are most far-reaching. 

This lecture is based on Polchinski's original paper~\cite{Joe1}
that recognized D-branes as genuine non-perturbative excitations of string
theory. It is not a review, but rather a concise foreword to the
already large literature on the many different  facets of the subject:
duality checks, black-hole entropy, D-brane scattering and
D-brane probes, orientifold compactifications. Some of these topics
 are covered in other recent reviews
~\cite{Joe2,John,Michael,Paul,Jan,Duff,K33,BHrev,Igor}, but the story is
 unfolding fast and it is still unclear where
 it will take us.
\eject

\section{Ramond-Ramond fields}

\subsection{Chiral Bispinors}
\label{subsec:bispinor}

  With the exception of the heterotic string, all other consistent
string theories  contain in their spectrum antisymmetric tensor
fields  coming from the  Ramond-Ramond sector.
 This is the case for the
type-IIa and IIb superstrings, as well as for the type-I theory whose
closed-string states are a subset of those of type-IIb.
 The spectrum of a type-II theory is obtained as
 a tensor product of a left- and a right-moving world-sheet
sector, each of which contains at the massless level a
10d  vector and a 10d  Weyl-Majorana spinor. This is
depicted figuratively as follows:
$$
\Bigl( \vert \mu\rangle  \oplus \vert\alpha \rangle \Bigr)_{left}
 \ \otimes \ 
\Bigl( \vert \nu\rangle  \oplus \vert\beta \rangle \Bigr)_{right}
$$
where
 $\mu,\nu = 0,... ,9$ and $\alpha,\beta = 1,... ,16$
 are
respectively vector and   spinor indices.
Bosonic fields thus include a two-index tensor,  which can be
decomposed into  symmetric traceless,  trace,  and antisymmetric
parts: these are the usual fluctuations of the  graviton
 ($G_{\mu\nu}$), dilaton ($\Phi$) and Neveu-Schwarz Neveu-Schwarz
antisymmetric tensor ($B_{\mu\nu}$). In addition massless bosonic
fields include a Ramond-Ramond bispinor $F_{\alpha\beta}$,
defined as the polarization in the corresponding
vertex operator
\begin{equation}
V_{RR} \sim  F_{\alpha\beta}(p) \int d^2\zeta\ 
S^\alpha (i\Gamma^0{\bar S})^\beta e^{ipX}
\label{eq:vertex}
\end{equation}
Here
 $S$, ${\bar S}$ are the left and right fermionic emission
vertices,  $p$ the 10d momentum, and the inclusion of the
$\Gamma^0$ ensures that under a Lorentz transformation
$F$ transforms by a similarity transformation.
Such bispinors can be decomposed in
a complete basis of all gamma-matrix
antisymmetric products
\footnote{I follow the conventions of  Green, Schwarz and Witten
~\cite{GSW} : the 10d gamma matrices are purely imaginary
and obey the algebra $\{\Gamma^\mu, \Gamma^\nu\} =-2\eta^{\mu\nu}$
with metric signature $(-+...+)$.
Factors of $i$ are  inserted appropriately so that
the bispinor and its tensor components are all real.
The chirality operator is
 $\Gamma_{11} = \Gamma_0\Gamma_1...\Gamma_9$, and the Levi-Civita
tensor $\epsilon^{01...9}= 1$.
}
\begin{equation}
 F_{\alpha\beta} = \sum_{k=0}^{10} {i^{k}\over k!}
  F_{\mu_1...\mu_k}
(\Gamma^{\mu_1...\mu_k})_{\alpha\beta}
\end{equation}
where
\begin{equation}
 \Gamma^{\mu_1...\mu_k} \equiv {1\over k!} \Gamma^{[\mu_1}...
\Gamma^{\mu_k]}
\end{equation}
and the $k=0$ term stands by convention for the
identity in spinor space. As a result the Ramond
Ramond massless fields are a collection of antisymmetric
Lorentz tensors.

 These tensor
components are not all independent, because the bispinor
field has definite chirality projections
\begin{equation}
\Gamma_{11} F = \pm  F \Gamma_{11} = F
\label{eq:chiral}
\end{equation}
The choice of sign distinguishes between the type-IIa and
type-IIb models, for which the two spinors are, respectively,
of opposite or same chirality in ten dimensions.
Taking into account our definition of the bispinor field, we see
that   type-IIa corresponds to the sign plus and 
type-IIb to the sign minus. To rewrite eqs.(~\ref{eq:chiral})
 in terms of
the tensor fields we need
  the gamma identities
\bea
\Gamma_{11} \Gamma^{\mu_1...\mu_k} &=&
{(-)^{[{k\over 2}]} \over (10-k)!}
 \epsilon^{\mu_1...\mu_{10}} \Gamma_{\mu_{k+1}...\mu_{10}}
\\
\Gamma^{\mu_1...\mu_k} \Gamma_{11} &=&
 {(-)^{[{k+1\over 2}]}\over (10-k)!}
 \epsilon^{\mu_1...\mu_{10}} \Gamma_{\mu_{k+1}...\mu_{10}}
\eea
with $[x]$ denoting the integer part of $x$. It follows easily
that only even-$k$ (odd-$k$) terms are allowed in the type-IIa
(type-IIb) case, and that furthermore all antisymmetric fields
obey the  duality relations
\begin{equation}
F^{\mu_1...\mu_k} = {(-)^{[{k+1\over 2}]}\over (10-k)!}
 \epsilon^{\mu_1...\mu_{10}} F_{\mu_{k+1}...\mu_{10}}
\label{eq:dual}
\end{equation}
We write these relations in short-hand form as $F_{(k)} =
 \pm
\ ^*F_{(10-k)}$. As a check note that the type-IIa theory
has  independent tensors with $k=0,2$ and $4$ indices, while
the type-IIb theory has $k=1,3$ and a self-dual $k=5$ tensor.
The number of independent tensor components adds up in both
cases to $16\times 16 = 256$:
\bea
& {\underline {\rm type-IIa:}}&\ \ \
1 + {10\times 9 \over 2!} + {10\times 9 \times 8 \times 7 \over
 4!}
 = 256\ , \nonumber \\
& & \nonumber \\
& {\underline {\rm type-IIb:}}& \ \ \
10 + {10\times 9 \times 8 \over 3!} + {10\times 9 \times 8
\times 7 \times 6 \over 2 \times  5!} = 256\ , \nonumber
\eea
which is  precisely  the number of components of a bispinor.

\subsection{Free Field Equations}
\label{subsec:freeeqs}

  The mass-shell or super-Virasoro
conditions for the vertex operator (~\ref{eq:vertex})
imply that the bispinor field obeys two
 massless Dirac equations
\begin{equation}
 (p_\mu \Gamma^\mu) F = F (p_\mu \Gamma^\mu) = 0 \ .
\end{equation}
To convert these to equations for the tensors we use the gamma
identities
\begin{equation}
\Gamma^\mu \Gamma^{\nu_1...\nu_k} =  \Gamma^{\mu\nu_1...\nu_k}
- {1\over (k-1)!} \eta^{\mu [\nu_1} \Gamma^{\nu_2...\nu_k]}
\end{equation}
\begin{equation}
 \Gamma^{\nu_1...\nu_k} \Gamma^\mu =  \Gamma^{ \nu_1...\nu_k\mu}
- {1\over (k-1)!} \eta^{\mu [\nu_k} \Gamma^{\nu_1...\nu_{k-1}]}
\end{equation}
with
 square brackets denoting the  alternating sum
over all permutations of the enclosed indices. After some
straightforward algebra one finds
\begin{equation}
 p^{[\mu} F^{\nu_1...\nu_k]} = p_\mu F^{\mu \nu_2...\nu_k} =
 0
\end{equation}
which are the Bianchi identity and free massless equation for
an antisymmetric tensor field strength. We may write these in
economic form as
\begin{equation}
 d F = d \ ^*F = 0
\end{equation} 
Solving the Bianchi identity locally
allows us to express the  $k$-index field strength as the
exterior derivative of a $(k-1)$-form potential
\begin{equation}
F_{\mu_1...\mu_k} =
{1\over (k-1)!} \partial_{[\mu_1} C_{\mu_2...\mu_k]}  ,
\end{equation}
or in short-hand notation
\begin{equation}
 F_{(k)} = d C_{(k-1)} \ .
\end{equation}
Thus the type-IIa theory has a vector ($C^\mu$) and a three-index
tensor potential ($C^{\mu\nu\rho}$) , in addition to a constant
non-propagating zero-form field strength ($F$), while the
type-IIb theory has a zero-form ($C$), a two-form ($C^{\mu\nu}$)
and a four-form potential ($C^{\mu\nu\rho\sigma}$), the latter
 with self-dual field strength.
 The number of physical transverse
degrees of freedom adds up in both cases to $64=8\times 8$:
\bea
& {\underline {\rm type-IIa:}}& \ \ \
8 + {8\times 7\times 6 \over 3!}= 64\ , \nonumber
\\& & \nonumber  \\
& {\underline {\rm type-IIb:}}& \ \ \
1 + {8\times 7  \over 2!} + {8\times 7 \times 6
\times 5   \over 2 \times  4!} = 64\ , \nonumber
\eea
which is precisely  the number of physical components of a bispinor.

\subsection{No RR  Charges in Perturbation Theory}
\label{subsec:nocharge}

A $(p+1)$-form potential couples naturally to a $p$-brane,
i.e. an extended object with $p$ spatial dimensions.
The coupling is the integral of the form over the
 $(p+1)$-dimensional world-volume of the brane:
\begin{equation}
\int_{world\atop vol}  C_{(p+1)} \equiv
\int d^{p+1}\zeta \ C_{\mu_1...\mu_{p+1}}(X)\  \partial_{1}
X^{\mu_1}
... \partial_{{p+1}}
X^{\mu_{p+1}} 
\end{equation}
where $X^\mu(\zeta^\alpha)$ are the target-space coordinates of the
brane \footnote{Lower-case Greek letters label both the
world-volume coordinates and the space-time spinors. 
Put in context, this should hopefully cause no confusion.}.
 For example  the elementary charge for 
  a vector potential is a point-particle or  0-brane, that for a
two-index tensor is a 1-brane or string,  and so on down the line.
In view of the duality  relation (~\ref{eq:dual}), 
  $p$-branes
and $(6-p)$-branes are  electric-magnetic duals of each
 other in ten dimensions.

 Now string perturbation
theory contains no such elementary RR charges: to start with
it has no branes other than the fundamental strings, which
could at most couple to the RR two-index tensor. That even this
doesn't happen follows from the fact that a trilinear coupling
$<s\vert C_{(k)}\vert s>$,  corresponding to the emission of a RR
 $k$-form from
any given string state, would violate  the conservation laws of separate
left and right fermion numbers, which are
valid to all orders in the genus expansion.
Note also that the RR vertex operator involves
the field strength rather than the potential, and that the field
 equations
and Bianchi identities entered in a completely symmetric way: this
duality of perturbation theory would have been
 destroyed by the presence
 of either electric or magnetic charges.

Most non-pertubative dualities require, on the other hand, the
existence
of such elementary charges. Two of these dualities involve the
 strong-coupling
limits of the type-IIa and type-IIb theories in ten 
dimensions~\cite{Dual,Duall}.
The former is conjectured to be ${\cal M}$ theory,
 whose low-energy limit is
11d supergravity. Indeed, the massless RR and NSNS fields
of the IIa theory can be obtained by dimensional reduction from eleven
dimensions:
\begin{equation}
G_{MN} \rightarrow G_{\mu\nu}, C_\mu, \Phi ; \ \ \
C_{MNR} \rightarrow C_{\mu\nu\rho}, B_{\mu\nu}
\end{equation}
with $M,N,R= 0,...,10$. 
In 11d supergravity there are however also Kaluza-Klein states
carrying charge
 under the off-diagonal metric components $A_\mu$.
Since this is a RR field in type-IIa  theory,
duality requires  the existence
of non-perturbative 0-brane charges.
Likewise the type-IIb theory has a SL(2,Z)
self-duality under which the dilaton and RR scalar $C$
transform together  as a complex modulus,
while  the two 2-index antisymmetric forms
$(B_{\mu\nu}, C_{\mu\nu})$ transform  as a doublet~\cite{SL2Z,noncom}.
 Since fundamental strings couple as elementary charges to
 $B_{\mu\nu}$, this duality requires  the existence of
non-perturbative  1-branes that are likewise 
charged  under $C_{\mu\nu}$.

Higher p-branes  fit similarly  in the  conjectured web of dualities.
 For instance type-IIa theory compactified to six dimensions
 on a $K_3$ surface is expected to be dual to the heterotic
 string compactified on a
four-torus~\cite{Dual,Duall}. The latter has extended gauge
 symmetry at special
 points of the
Narain moduli space. On the type-IIa side these charged massless
gauge bosons can only come from 2-branes
 wrapping around shrinking 2-cycles  of
 the K3 surface~\cite{conifold,K3,K33}.
A similar phenomenon occurs for Calabi-Yau compactifications:
the effective low-energy Lagrangian of Ramond-Ramond fields
has a logarithmic singularity at special (conifold) points
in the Calabi-Yau moduli space. This can be understood as due
to 3-branes  wrapping  around shrinking 3-cycles, and thus
becoming massless at these special points ~\cite{conifold,CY}.

 Early efforts consisted in describing all these exotic
states as (singular) solutions of the effective low-energy 
 supergravity~\cite{Duff}.
Following Polchinski's work it has become, however,  clear that
 they admit
a much simpler, exact and unified description
as allowed   endpoints for open strings, or D-branes.

\section{D(irichlet) branes}

\subsection{Definition}
\label{subsec:definition}

The bosonic part of the world-sheet action for a  free  string, in
flat space-time and in the conformal
gauge,  reads
\begin{equation}
S_{2d} =  \int_{\cal M} {d^2\zeta \over 4\pi\alpha^\prime}
 \ \partial_\alpha X^\mu
\partial^\alpha X_\mu
\end{equation}
with ${\cal M}$ some generic surface with boundary. For its variation
\begin{equation}
\delta  S_{2d} =  -\int_{\cal M} {d^2\zeta\over 2\pi\alpha^\prime}
 \ \delta X^\mu
\partial_\alpha\partial^\alpha X_\mu
+ \int_{\partial \cal M} {d\zeta^\alpha \over 2\pi\alpha^\prime}
 \ \delta X^\mu
\varepsilon_{\alpha\beta} \partial^\beta X_\mu
\end{equation}
to vanish,  the $X^\mu$ must be harmonic functions on the world sheet,
 and either of the following two conditions must hold on the boundary
\bea
 \partial_\perp X_\mu =& 0 & \ \ {(\rm Neumann)},\\  
{\rm  or} \ \ \  
 \delta X^\mu = &0 &\ \ {(\rm Dirichlet)} .
\eea
Neumann
 conditions  respect   Poincar\'e invariance  and
are hence momentum-conserving.
  Dirichlet conditions on the other hand
represent defects in space-time. They were studied in the past
in various guises, for instance  as
 sources for partonic behaviour in string theory~\cite{Green1},
 as heavy-quark
endpoints~\cite{quarks}, and as T-dual forms of open-string 
 compactifications~\cite{T,orientifold}, but  their  status of  
 non-perturbative states of string theory
 was not fully appreciated  in these early studies.

A static extended defect with $p$ spatial dimensions is described
 by the
boundary conditions
\begin{equation}
 \partial_\perp X^{0,1,..,p} = X^{p+1,..,9} = 0 \ 
, \label{eq:plane}
\end{equation}
that force open strings to move on a $(p+1)$-dimensional (world-volume)
hyperplane.  Since open strings
  do not propagate in the bulk in type-II
theory, their presence is intimately-tied to the existence of the
 defect
which we will refer to as D(p)-brane.
The D(p)-brane  is characterized by a   tension $T_{(p)}$,
and  charge density  under the Ramond-Ramond 
 $(p+1)$-form $\mu_{(p)}$,
defined through the effective world-volume action
\begin{equation}
 S_{\rm world\atop vol} =  T_{(p)} \int d^{p+1}\zeta \ e^{-\Phi/2}
\sqrt{\vert det{\hat G}_{\alpha\beta}\vert }
+ \mu_{(p)} \int d^{p+1}\zeta\ 
C_{(p+1)} \ ,
\end{equation}
where
\begin{equation}
{\hat G}_{\alpha\beta}= G^{\mu\nu}\partial_\alpha X_\mu
\partial_\beta X_\nu
\end{equation}
is the induced world-volume metric.
The   massless closed
strings coupling  to the D-brane have themselves a bulk action~\cite{GSW}
\begin{equation}
S_{\rm bulk} = -{1\over 2\kappa_{(10)}^2}
 \int d^{10}x \sqrt{-G} \Biggl[
 e^{-\Phi}
\Bigl( R- d\Phi^2  +{1\over 12} dB^2
\Bigr)
+\sum {1 \over 2 k!}
 F_{(k)}^2  \Biggr]
\label{eq:field}
\end{equation}
Here $k=0,2,4$ for type-IIa theory, $k=1,3$ for type-IIb, while
for the self-dual $F_{(5)}$ there is no covariant action we
may write down. Notice that  the
 bulk  Lagrangian coming  from the  sphere diagram 
is  multiplied by the usual factor
$e^{-\Phi}$, while the world-volume Lagrangian coming 
from  the disk diagram is  multiplied by $e^{-\Phi/2}$. 
These factors were absorbed in the Ramond-Ramond fields
through a rescaling
\begin{equation}
   C_{(p+1)} \rightarrow e^{\Phi/2} C_{(p+1)} \ .
\end{equation}
A carefull analysis  shows indeed that it is the field strength
 of rescaled
potentials  that satisfies the usual Bianchi identity and
  Maxwell equation  when
the dilaton varies  ~\cite{rescaling}. 

We will now  calculate the values of $T_{(p)}$
 and $\mu_{(p)}$ and check that they are compatible with supersymmetry
 and with the
(minimal) charge-quantization condition.
The charge density and tension could  be extracted in principle from
 one-point
functions on the disk. Following Polchinski ~\cite{Joe1}
 we will however prefer
to extract them from the interaction energy between two 
parallel identical D-branes.
This way of thinking avoids  the technicalities of 
normalizing vertex operators correctly,
and extends naturally to the study of D-brane dynamics~\cite{me,dyn}.

\subsection{Static Force: Field-Theory Calculation}
\label{subsec:force2}

Viewed as solitons of 10d supergravity, two D-branes interact
by exchanging gravitons, dilatons and antisymmetric tensors.
This is a good approximation, provided their  separation $r$
is large compared to the string scale,
in which case the effective actions, eqs. (3.6-8) can be trusted.
To decouple the propagators of the graviton and dilaton, 
we pass  to  the Einstein metric
\begin{equation}
g_{\mu\nu} = e^{-\Phi/4}  G_{\mu\nu} \ ,
\end{equation}
in terms of which the effective  actions take the form
\bea
S_{bulk}= -{1\over 2\kappa_{(10)}^2}
& \int& d^{10}x \sqrt{-g}\ \Bigl[
 R + {1\over 8} d\Phi^2 
 +{1\over 12} e^{-\Phi/2} dB^2 \nonumber \\
& +& \sum  {1\over 2(p+2)!}
e^{(3-p)\Phi/4} d C_{(p+1)}^2 \ \Bigr]
\eea
and
\begin{equation}
 S_{world\atop vol} =  T_{(p)} \int d^{p+1}\zeta\ 
 e^{(p-3)\Phi/8}
\sqrt{\vert det{\hat g}_{\alpha\beta}\vert} +
  \mu_{(p)} \int d^{p+1}\zeta\   C_{(p+1)} \ . 
\end{equation}
To leading order in the coupling constant
the interaction energy comes from  the exchange of
 a single graviton, dilaton or Ramond-Ramond field, and  reads
\begin{equation}
{\cal E}(r){\cal T} =
- 2\kappa_{(10)}^2 \int d^{10}x \int d^{10}{\tilde x}
\ \Bigl[ 4 j_{\Phi} \Delta {\tilde j_{\Phi}}
-  j_{C} \Delta  {\tilde j_{C}} +
T_{\mu\nu} \Delta^{\mu\nu,\rho\tau}{\tilde T}_{\rho\tau}
\Bigr]
\end{equation}
Here $j_{\Phi}$, $j_C$ and $T_{\mu\nu}$ are the sources for
the dilaton, RR form and graviton obtained by linearizing
the world-volume action for one of the branes, while the
tilde quantities refer to the other.
  $\Delta$ 
and $\Delta^{\mu\nu,\rho\tau}$
are the scalar 
and  the graviton propagators in $10$
dimensions, evaluated at
the  argument ($x-{\tilde x}$), and ${\cal T}$ the total interaction
time.
To simplify notation, and since only one component of
$C_{(p+1)}$ couples to a  static planar D(p)-brane, we have
 dropped the obvious tensor structure of the RR field.

For such a static planar defect the sources take the simple form
\bea
j_{\Phi} &=& {p-3\over 8} T_{(p)} \delta(x^{\perp}) \nonumber \\
j_{C}& =& \mu_{(p)}  \delta(x^{\perp}) 
\label{eq:sources}
\eea
\begin{equation}
T_{\mu\nu} =  {1\over 2}  T_{(p)}   \delta(x^{\perp}) \times
\cases{ &$\eta_{\mu\nu}$
\ {\rm if}\ \ $\mu,\nu \leq p$ \cr
& $0$ \ \ \ \ \  {\rm otherwise} \cr}
\end{equation}
where
the  $\delta$-function localizes  the defect in transverse space.
The graviton propagator 
 in the De Donder gauge and in $d$ dimensions reads
~\cite{DeDonder}
\begin{equation}
 \Delta_{(d)}^{\mu\nu,\rho\tau} = (\eta^{\mu\rho}\eta^{\nu\tau}
+ \eta^{\mu\tau}\eta^{\nu\rho} -\frac{2}{d-2}
\eta^{\mu\nu}\eta^{\rho\tau} )\Delta_{(d)} \ ,
\end{equation}
where
\begin{equation}
\Delta_{(d)}(x) = \int {d^{d} p\over (2\pi)^{d}}
{e^{ipx} \over p^2} \ .
\end{equation}
Putting all this together and  doing some straightforward algebra
we obtain
\begin{equation}
{\cal E}(r) = 2 V_{(p)} \kappa_{(10)}^2
\  [ \mu_{(p)}^2 -  T_{(p)}^2 ]\ 
\Delta_{(9-p)}^E(r)\ ,
\label{eq:sta}
\end{equation} 
where $V_{(p)}$ is the (regularized)  p-brane volume and 
$\Delta_{(9-p)}^E(r)$ is the (Euclidean) scalar
 propagator in $(9-p)$ transverse dimensions. The net force
is as should be expected the difference between RR repulsion and
gravitational plus dilaton attraction.

\begin{figure}
%
\centerline{\psfig{figure=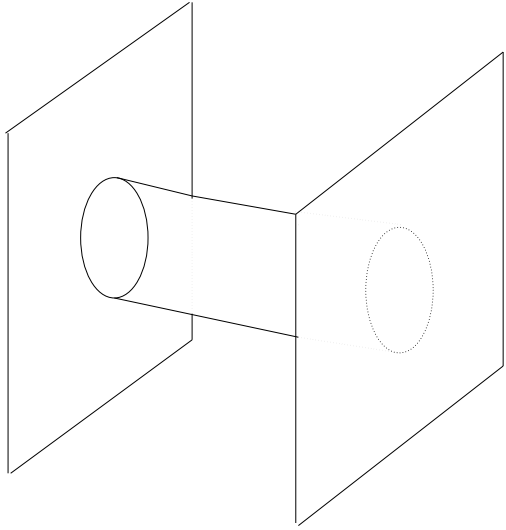,height=7cm}}
%
%
\caption{Two D-branes interacting through  the exchange of
a closed string. The diagram has a dual interpretation as
Casimir force due to vacuum fluctuations of open strings.
\label{fig:radish}}
\end{figure}

\subsection{Static Force: String Calculation}
\label{subsec:force1}

The exchange of all closed-string modes, including the massless
  graviton, dilaton and  RR $(p+1)$-form,
is given  by the  cylinder diagram of figure  1. 
Considered, however,  as an annulus, this same diagram also admits a dual
and, from the field-theory point of view,  surprising
interpretation: 
the  two D-branes interact by modifying the
vacuum fluctuations of (stretched) open strings, in the same  way that
two superconducting
plates attract by  modifying  the vacuum fluctuations
of the photon field. It is this simple-minded duality which
may. as we will see below,  revolutionize our thinking about space-time.

The one-loop vacuum energy  of  open strings reads
\bea
& &{\cal E}(r)  =
-{V^{(p)}\over 2} 
 \int {d^{p+1}k\over (2\pi)^{p+1}}\ \int_0^\infty
{dt\over t}\  {\rm Str} e^{-\pi t(k^2+ M^2)/2}=
 \nonumber \\
& & \label{eq:static}\\
&=&  - 2\times {V^{(p)} \over 2}\int_0^\infty
{dt\over t} (2\pi^2 t)^{-(p+1)/2} e^{-r^2 t/2\pi}\ Z(t)
 \ , \nonumber
\eea
where 
\begin{equation}
Z(t) = {1\over 2} \sum_{s=2,3,4} (-)^s \theta^4_s(it/2)
 \eta^{-12}(it/2)
\end{equation}
is the usual spin structure sum obtained by supertracing over
open-string oscillator states, and we have set $\alpha^\prime =
1/2$. Strings stretching between the two D-branes
  have at the $N$th oscillator level a mass
 $M^2= (r/\pi)^2 + 2N$, so that their  vacuum fluctuations are 
modified when we separate the D-branes. The vacuum energy
of open strings
 with both endpoints on the same  defect is, on the other hand,
$r$-independent and has been omitted. 
Notice also the (important)
 factor of $2$ in front of the second line: it 
accounts for  the two possible orientations of the
 stretched  string,

 The first remark concerning  the above expression,
is that it vanishes by the well-known  $\theta$-function  identity. 
Comparing with eq. (~\ref{eq:sta}) we conclude that
\begin{equation}
T_{(p)} = \mu_{(p)} \ ,
\label{eq:BPS}
\end{equation}
so that RR repulsion cancels exactly the gravitational and
dilaton attraction. This is a consequence of 
space-time supersymmetry: the relation  (~\ref{eq:BPS})
between
tension and charge is in fact the BPS condition for short
multiplets of the N=8 supersymmetry algebra  in ten dimensions.
Put differently,  Dirichlet conditions corresponding to
a single static defect
identify left- and right-moving
spin fields on the boundary: half of the target-space supersymmetries
are thus still linearly realized in perturbative string theory,
i.e. they are  unbroken in the presence of the D-brane. 
As is usual for
 BPS states,  identical and parallel D-branes 
 exert no net force on one another. 
This is analogous
to the cancellation of Coulomb repulsion and Higgs-scalar
attraction between the 'tHooft-Polyakov monopoles of
N=4 supersymmetric Yang-Mills.

   To extract the actual value of $T_{(p)}$ we must  
separate in the diagram the exchange of RR and NS-NS closed-string
states. These are characterized by world-sheet fermions
which are periodic, respectively antiperiodic around the
cylinder, so that they  correspond to the $s=4$, respectively
$s=2,3$ open-string spin structures. In the large-separation
limit ($r\to\infty$) we may furthermore expand the integrand
near 
$t\sim  0$:
\begin{equation}
Z(t) \simeq (8-8)\times \left(\frac{t}{2}\right)^4 + o(e^{-1/t}) \ .
\end{equation}
Using the
 integral representation
\begin{equation}
\Delta^E_{(d)}(r) = {\pi\over 2} \int_0^\infty dl
(2\pi^2 l)^{-d/2} e^{-r^2/2\pi l} \ .
\end{equation}
 and restoring correct mass units we obtain 
\begin{equation}
{\cal E}(r) = V^{(p)} (1-1) 2\pi (4\pi^2\alpha^\prime)^{3-p}
\Delta_{(9-p)}^E(r)\  +\  o(e^{-r/\sqrt{\alpha^\prime}}) \ .
\end{equation}
Comparing with the field-theory calculation we can finally extract
the tension and charge-density of  D(p)-branes ,
\begin{equation}
T_{(p)}^2 = \mu_{(p)}^2 = \frac{\pi}{\kappa_{(10)}^2}
(4\pi^2\alpha^\prime)^{3-p}\ .
\end{equation}
These are  fixed as should be expected 
for  intrinsic excitations of a  fundamental theory. We
will now see that they also pass the first non-trivial
consistency condition.

\begin{figure}
\centerline{\psfig{figure=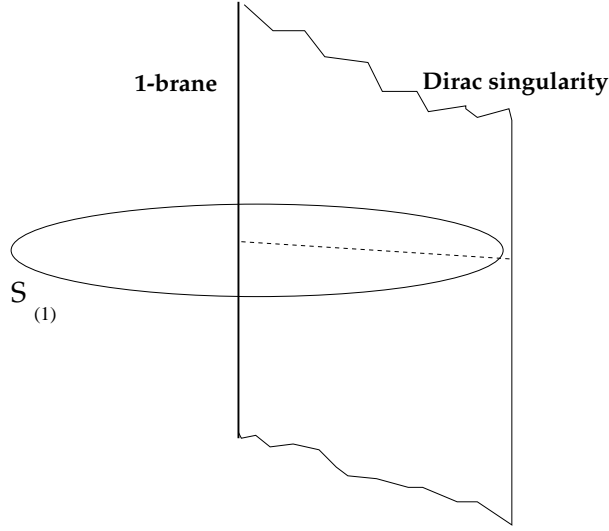,height=7cm}}
\caption{A 1-brane  creates
a 3-index ``electric'' field $F_{(3)}$. Electric flux
in d=4 space-time dimensions  is given
by an integral of the dual vector over a 1-sphere.
 The magnetic potential is  a scalar field
with a  discontinuity across the depicted  Dirac
sheet singularity.
\label{fig:ra}}
\end{figure}

\subsection{Charge Quantization}
\label{subsec:quantization}

Dirac's quantization condition for electric and magnetic charge
has an  analog for extended objects in higher
dimensions ~\cite{DNT}.
Consider a D(p)-brane sitting at the origin, and 
integrate  the field equation that follows from the action
(3.6-8) over  the transverse space. Using Stoke's theorem,  one finds
\begin{equation}
\int_{S_{(8-p)}} *F_{(p+2)} = 2\kappa_{(10)}^2  \mu_{(p)}
\end{equation}
where ${S_{(8-p)}}$ is a sphere around the defect as illustrated
 in figure 2. 
This is the analog of  Gauss' law.
Now 
 Poincar\'e duality tells us  that
\begin{equation}
*F_{(p+2)} = \pm F_{(8-p)}  \simeq \pm d C_{(7-p)} \ ,
\end{equation}
where  the   potential $C_{(7-p)}$ is not globally defined
since the p-brane is a source in the
Bianchi identity for $F_{(8-p)}$.
Following Dirac we may define a smooth potential everywhere
 except along  a singular (hyper)string cutting
$S_{(8-p)}$ on a hypersphere $S_{(7-p)}$.
This singularity  is dangerous  since a Bohm-Aharonov 
experiment involving (6-p)-branes might detect it. 
Indeed, the wave-function of
a  (6-p)-brane transported 
around the  singuality  picks a phase
\bea
{\rm Phase}=&& \mu_{(6-p)} \int_{S_{(7-p)}} C_{(7-p)}= \nonumber \\
&=& \mu_{(6-p)} \int_{S_{(8-p)}} F_{(8-p)} =\ \pm  2\kappa_{(10)}^2
\mu_{(p)}\mu_{(6-p)} \ . \label{eq:Dirac}
\eea
For the (hyper)string not to be observed we
must have
\begin{equation}
{\rm Phase}= 2\pi n \ .
\end{equation}   
The D-brane charges found in the previous section satisfy this
condition with $n=1$!
The simple annulus
diagram somehow knows about this non-perturbative consistency check
~\cite{Joe1}. Furthermore we have just learned that
 D-branes are the minimal RR charges
allowed in the theory, so one may conjecture  that there
are no others.

\section{ Gauge Theory and Geometry}

\subsection{Non-commutative Space-time}
\label{subsec:Noncom}

Although the  full dynamics of  solitons cannot be separated from the
field theory in which they belong, their low-energy dynamics
can be approximated by quantum mechanics in the moduli-space
of zero modes. For an extended p-brane defect,  the zero modes
give rise to massless world-volume fields  and the quantum mechanics
becomes a (p+1)-dimensional field theory.
Similar considerations apply to a D(p)-brane: its perturbative
excitations are described by a 
 full-fledged 
 open string theory, whose low-energy limit is an
 abelian supersymmetric Yang-Mills, 
dimensionnally-reduced from ten down to  (p+1) dimensions,
\begin{equation}
A^\mu \rightarrow  A^\alpha (X^\alpha) , A^i(X^\alpha) \ .
\label{eq:reduction}
\end{equation}
I am using  here a  physical parametrization  
 $X^{\alpha} = \zeta^{\alpha}$
($\alpha=0,..,p$), corresponding to
  a planar static background defect like
the one of  eqs. (3.5).  The world-volume scalars are proportional
 to the transverse coordinates of the brane
\begin{equation}
X^i  = 2\pi\alpha^\prime A^ i  \ \ \ \ 
\ (i=p+1,..,9).
\end{equation}
They are the Goldstone modes of broken translation invariance,
forced by  supersymmetry
 to be part of an entire vector  multiplet. 

\begin{figure}
\centerline{\psfig{figure=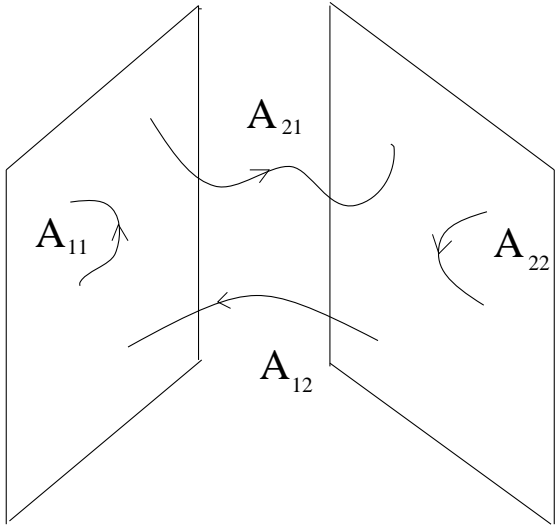,height=7cm}}
\caption{
A  D-brane sandwitch, and the
 four types of open strings giving rise to massless states
 in the coincidence limit. 
\label{fig:nonabelian}}
\end{figure}

 That much can  be in fact deduced within the context of
N=2 supergravity in ten dimensions~\cite{Duff}.
 Where the role of string theory
 becomes important  is in the presence of  more than one
 D-branes.
In addition to the massless vector multiplets describing the
positions of each defect, there are now extra potentially-light fields
corresponding to the open strings that stretch between the 
D-branes. These are the strings responsible for the
Casimir force of the previous chapter. 
In the simplest case
of two parallel  D(p)-branes (figure 3), the ensuing low-energy field theory
is a dimensionnally-reduced 
 supersymmetric Yang-Mills with
non-abelian gauge group~\cite{noncom}
\begin{equation}
U(2) \simeq U(1)_{cm} \times SU(2)_{relative} \ .
\end{equation}
This is indeed the low-energy limit of an oriented open-string
theory with a Chan-Patton index $i=1,2$ labelling the two possible string
endpoints. The 
 abelian vector multiplet $A_{cm}= (A_{11}+A_{22})/\sqrt{2}$
describes the dynamics of the center of mass, while the
non-abelian  SU(2) describes the reduced relative motion.
At non-zero separation
\begin{equation}
r= 2\pi\alpha^\prime \vert \langle A_{11}^i -
 A_{22}^i \rangle\vert
,
\end{equation}
the non-abelian theory is in the spontaneously-broken phase:
charged gauge bosons are stretched open strings that have  a mass
$M = r/2\pi\alpha^\prime$.

  Figure 3 summarizes in itself much of the  new insights brought
by D-branes. In a nutshell it teaches us that the short-distance
structure of space-time is intimately related to supersymmetric
gauge field theory. 
The details of the theory depend of course  on the precise 
 configuration of D-branes, but the mapping between ultraviolet
gravity and infrared gauge theory seems to be 
universally valid~\cite{IR}.
This fact underlies the
successfull counting of the Beckenstein-Hawking entropy
for extremal (BPS) configurations of D-branes~\cite{BH,BHrev}.
It is also the essential ingredient of  the conjectured equivalence
between infinite-momentum {\cal M} theory, and
M(atrix)-model quantum mechanics~\cite{Matrix}.

\subsection{T-duality}
\label{subsec:T}

 The dimensional reduction (~\ref{eq:reduction}) suggests that
the transverse D-brane coordinates are gauge fields in some
invisible internal dimensions. The precise correspondence involves a
(perturbative) T-duality transformation, which we will now
make more explicit. T-duality is a
  local symmetry  of the fundamental  theory,
 that transforms  both the background fields and the 
 string excitations around them~\cite{GPR}.
 Recall that
a closed-string coordinate along a compact (ninth)  dimension
reads 
\begin{equation}
X^9(z,\bar z) = X^9(z) +{\tilde X}^9(\bar z)\ ,
\end{equation}
 where
\bea
X^9(z) = x^9 - {i\over 2} \left( \frac{n_9}{2R_9} +
 m_9R_9\right){\rm ln}z
 + {i\over 2} \sum_{k\not= 0} {a^9_k\over k} z^{-k}\nonumber  \\
{\tilde X}^9(\bar z) = {\tilde x}^9 -
 {i\over 2} \left( \frac{n_9}{2R_9} -  m_9R_9\right)
{\rm ln}{\bar z}
 + {i\over 2} \sum_{k\not= 0} {{\tilde a}^9_k\over k} {\bar z}^{-k}
\label{eq:modes}
\eea
Here $n_9$ and $m_9$ are the (integer in appropriate units)
momentum and winding,
and we are using radial world-sheet coordinates 
$z= e^{\tau+i\sigma}$. The simplest T-duality transformation
inverts  the radius, and interchanges  winding with momentum:
\begin{equation}
{\tilde X}^{9 \prime} =  -{\tilde X}^9 \ , \ \ 
 R_9^\prime = \frac{1}{2R_9}\  
\ {\rm and}\ \ ( n_9^\prime, m_9^\prime )= (m_9, n_9)\ .
\end{equation}
It is a hybrid parity operation, whose action
is restricted  to  the antiholomorphic world-sheet sector.

The parity operator in spinor space is  $-i\Gamma^9\Gamma_{11}$, 
so  bispinor fields will transform as follows:
\begin{equation}
F^\prime = -i F \Gamma^9\Gamma_{11}  \ .
\end{equation}
Using the $\Gamma$-matrix identities of chapter 2, we may rewrite
this relation in  component form, 
\bea
F^\prime_{\mu_1 ... \mu_k} = - F_{9 \mu_1 ... \mu_k} \nonumber \\
F^\prime_{9\mu_1 ... \mu_k} = F_{\mu_1 ... \mu_k}
\eea
where $\mu_i \not =9$ for all $i$. The duality exchanges even-k  with
odd-k antisymmetric forms, and hence also type IIa with type IIb
backgrounds. Consistency requires that it also transform
even-p to odd-p branes and vice versa.

To see how this comes about let us look at a D(p+1)-brane
that wraps around the first $p$,  as well as around the 9th 
dimension. We concentrate on the 9th coordinate of the open
strings that live  on the D-brane.
Except for an extra factor 2 multiplying zero modes, because
open strings are  parametrized by $\sigma\in [0,\pi ]$,
the mode expansions  (~\ref{eq:modes}) stay  valid.
The world-sheet is  now the infinite strip, mapped
in radial coordinates to the upper half complex plane.
Imposing  appropriate
(Neumann) conditions on the boundary, 
\begin{equation}
X^9(z) = {\tilde X}^9(\bar z) \ \ {\rm at\   Im}z=0 \ ,
\end{equation}
identifies  left and right oscillators and sets the winding
number $m_9 = 0$. This is consistent with the fact that
open strings   move freely along the ninth dimension but
cannot wind.

Now the action of a T-duality transformation changes  the  Neumann
condition to Dirichlet,
\begin{equation}  X^{9\prime}(z) = -{\tilde X}^{9\prime}
(\bar z) \ \ {\rm at\   Im}z=0 \ ,  
\end{equation} 
and   momentum to winding~\cite{T}. The
(p+1)-brane  is thus transformed to a p-brane:
 indeed, open strings cannot
move along the 9th dimension any more,  but since their endpoints
are now attached they can wind. The inverse transformation is
also true: a 
 p-brane transverse to the 9th dimension becomes a (p+1)-brane
that wraps  around it. All this is compatible with the  
transformation of the RR forms to which the D-branes couple.
It is also compatible~\cite{Joe2}  with the tension formula, eq. (3.25),
if one takes into account the fact that  a T-duality
shifts the string coupling constant,
\begin{equation}
\Phi^\prime = \Phi+ {\rm ln}(\alpha^\prime/R^2) \ .
\end{equation}

 Since T-duality is an isomorphism  on  the algebra of vertex
operators, the full open string theory,  and its low-energy
limit  in particular are unchanged. The gauge fields, which  start
out as functions of $X^{0,..,p}$,  and of $X^9$ or equivalently
the  conjugate momentum $n_9$, become after the transformation
functions of the same first  (p+1) coordinates,  and of the dual
coordinate $X^{9\prime}$ or equivalently the winding number
$m_9^\prime$. The latter dependence can be ignored in the
$R_9\to 0$ or $R_9^\prime\to\infty$ limit: this is precisely the
limit of a dimensionally-reduced (p+1)-brane, or of a dual
p-brane transverse to  an  uncompactified ninth dimension.

\subsection{The Power of Lorentz Invariance}
\label{subsec:impl}

 To  illustrate the power of this duality~\cite{me,WZ,last}, 
consider the   0-brane of type-IIa theory performing some
one-dimensional motion
\be
X^0 = \tau  \ , \ \  X^1= f(\tau)\ .
\ee
Its  action, eq. (3.6), 
 in flat space-time but non-zero  background for the 
RR vector potential,  reads 
\be
S_{world\atop line} = 
 T_{(0)} \int d\tau \  \sqrt{1- {\dot f}^2}\  +\ 
\mu_{(0)}\int d\tau\  (C_{0} + {\dot f}  C_{1}) \  \ + ...
\ee
where the dots stand for acceleration terms that we neglected.
Let us now T-dualize the $X^1$ coordinate: this changes
  the 0-brane
to a 1-brane,  and  the  velocity to an electric field
\be
{\cal F}_{01} = \partial_0 A_1  =  {\dot f} /2\pi\alpha^\prime \ .
\ee
It also transforms  the RR background,
taken to be  $X^{0,1}$-independent for simplicity, as follows:
\be
C_1  \rightarrow  C \ ,\  \ \  C_{\mu} \rightarrow C_{\mu 1} \
\ \ (\mu\not= 1) .
\ee
The world-volume action in dual language  takes therefore the form
\bea
S_{world\atop vol} = 
 T_{(1)} &\int&  d^2\zeta\  \sqrt{1-
 (2\pi\alpha^\prime {\cal F}_{01})^2} +\\
&+& 
\mu_{(1)}\int d^2\zeta\ (C_{01} + 2\pi\alpha^\prime {\cal F}_{01} C)
\ + \ ...
\eea
in the physical parametrization  $\zeta^{0,1}= X^{0,1}$.

Something surprising  has happened here: from the simple 
point-particle Lagrangian,  fixed  completely
 by Poincare invariance, we are extracting seemingly non-trivial
information about  open-string gauge dynamics.
What we  are learning in particular is
 (i) that the effective Lagrangian
for slowly-varying electromagnetic fields must be of the
Born-Infeld type~\cite{BI}, and (ii) that there are  extra Wess-Zumino
terms~\cite{Li,WZ},  which when generalized appropriately 
include  the anomaly-cancelling Green-Schwarz couplings~\cite{GSW}. 
Both of these key features follow directly
from T-duality and  Lorentz invariance. They also follow presumably 
from the requirement that one of the two ten-dimensional
supersymmetries, broken spontaneously by the presence of the D-brane,
be realized  non-linearly
on the world volume~\cite{Bagger}.  I suspect these  two points of view
are related though I am not sure of the details. 
There is finally another intriguing facet to the argument:
the Born-Infeld action has a limiting electric field and a natural
dissipation mechanism~\cite{Mass} to account for
 it\footnote{To be sure,
near the limiting field there is no real quantitative
control of the dissipation rate.}.
In (T)-dual language
this would imply  that the speed of light is a dynamical rather than a
mere kinematic limit~\cite{me}, a conclusion befitting a fundamental
 theory of gravity.

I will stop here even though the fun is only  starting! The interested
reader will find many  more details in Polchinski's nice review
~\cite{Joe2}, and in the more than hundred papers on D-branes
in hep-th this past year.

\section*{Acknowledgments}
 I thank the organizers for the  invitation to speak in a very
stimulating conference, and aknowledge partial support from the
EEC grant CHRX-CT93-0340.

\section*{References}
\newcommand{\Journal}[4]{{#1} {\bf #2}, #3 (#4)}
\newcommand{\NCA}{\em Nuovo Cimento}
\newcommand{\NIM}{\em Nucl. Instrum. Methods}
\newcommand{\NIMA}{{\em Nucl. Instrum. Methods} A}
\newcommand{\NPB}{{\em Nucl. Phys.} B}
\newcommand{\PLB}{{\em Phys. Lett.}  B}
\newcommand{\PRL}{\em Phys. Rev. Lett.}
\newcommand{\PRD}{{\em Phys. Rev.} D}
\newcommand{\ZPC}{{\em Z. Phys.} C}

\end{document}